\documentclass[%
 reprint,
 amsmath,amssymb,
 aps,
]{revtex4-1}

\usepackage{tabularx}
\usepackage{booktabs}
\usepackage{upgreek}
\usepackage{nicefrac}
\usepackage{hyperref}
\usepackage{natbib}
\usepackage{xcolor}
\usepackage{graphicx}
\usepackage{dcolumn}
\usepackage{bm}

\begin{document}

\preprint{APS/123-QED}

\title{Transition Frequencies and Hyperfine Structure in $^{113,115}$In$^+$: Application of a Liquid-Metal Ion Source for Collinear Laser Spectroscopy}

\author{Kristian K\"onig}
 \altaffiliation[Current address: ]{National Superconducting Cyclotron Laboratory, Michigan State University, East Lansing, USA}
 \email{kkoenig@ikp.tu-darmstadt.de}
\affiliation{Institut f\"ur Kernphysik, Technische Universit\"at Darmstadt, Schlossgartenstr. 9, 64289 Darmstadt, Germany}

\author{J\"org Kr\"amer}
\affiliation{Institut f\"ur Kernphysik, Technische Universit\"at Darmstadt, Schlossgartenstr. 9, 64289 Darmstadt, Germany}

\author{Phillip Imgram}
\affiliation{Institut f\"ur Kernphysik, Technische Universit\"at Darmstadt, Schlossgartenstr. 9, 64289 Darmstadt, Germany}

\author{Bernhard Maa\ss}
\affiliation{Institut f\"ur Kernphysik, Technische Universit\"at Darmstadt, Schlossgartenstr. 9, 64289 Darmstadt, Germany}

\author{Wilfried N\"ortersh\"auser}
\affiliation{Institut f\"ur Kernphysik, Technische Universit\"at Darmstadt, Schlossgartenstr. 9, 64289 Darmstadt, Germany}

\author{Tim Ratajczyk}
\affiliation{Institut f\"ur Kernphysik, Technische Universit\"at Darmstadt, Schlossgartenstr. 9, 64289 Darmstadt, Germany}

\date{\today}

\begin{abstract}
We demonstrate the first application of a liquid-metal ion source for collinear laser spectroscopy in proof-of-principle measurements on naturally abundant In$^+$. The superior beam quality, i.e., the actively stabilized current and energy of a beam with very low transverse emittance, allowed us to perform precision spectroscopy on the $5s^2\;^1\mathrm{S}_0 \rightarrow 5s5p\;^3\mathrm{P}_1$ intercombination transition in $^{115}$In$^+$, which is to our knowledge the slowest transition measured with collinear fluorescence laser spectroscopy so far. By applying collinear and anticollinear spectroscopy, we improved the center-of-gravity frequency $\nu_\mathrm{cg}=1\,299\,617\,759.\,3\,(1.2)$ and the hyperfine constants $A=6957.19\,(28)$\,MHz and $B=-443.7\,(2.4)$\,MHz by more than two orders of magnitude. A similar accuracy was reached for $^{113}$In$^+$ in combination with literature data and the isotope shift between both naturally abundant isotopes was deduced to $\nu(^{113}\mathrm{In})-\nu(^{115}\mathrm{In})=696.3\,(3.1)$\,MHz.
Nuclear alignment induced by optical pumping in a preparation section of the ion beamline was demonstrated as a pump-and-probe approach to provide sharp features on top of the Doppler broadened resonance profile. 
\end{abstract}

\maketitle


\section{\label{sec:Introduction}Introduction}
Sources for atomic, ionic and molecular beams find applications in many fields of research. In this paper, we describe the first utilization of a liquid-metal ion source (LMIS) for collinear laser spectroscopy and present characterizing measurements using an In$^+$ beam. Due to the superior performance, such ion sources found a multitude of applications, ranging from focused-ion-beam microprobes over mask/circuit repair and manipulation to lithography (see e.g. \cite{Swanson.1983, Mair.1992, Orloff.2003, Bischoff.2005} and references therein). Due to their compact design, high reliability and low maintenance requirements, these sources are even used as ion thrusters for space crafts \cite{Tajmar.2009, Tajmar.2013}. They are able to deliver ion currents of up to several mA
, which allows to reduce the transverse emittance significantly by trimming the beam \cite{Koenig.2020}. 
Operation with a few pure elements is possible, among them especially indium and gallium, which have very low melting points and low vapor pressures, but there are many suitable eutectic alloys containing most metallic elements \cite{Bischoff.2016}. This also allows the production of molecular beams of various charge states.
While most metallic ions are also accessible with standard plasma ion sources like electron cyclotron resonance ion sources or Penning ion sources, these advantages and the relatively low cost make LMIS an interesting alternative.

We will report on the adaption of a liquid-metal ion source for the needs of collinear laser spectroscopy and present atomic spectra obtained from beams of natural abundant indium. Our realization actively stabilizes ion current and ion energy, leading to optimal conditions for these experiments. Furthermore, we demonstrate laser spectroscopy in the deep UV, which sets high demands, particularly on laser system and fluorescence detection, and present optical pumping schemes that allow spectroscopy on narrow spectral features to overcome precision limits due to broad resonance signals.

We chose In$^+$ for the characterization measurements since the $5s^2\;^1\mathrm{S}_0 \rightarrow 5s5p\;^3\mathrm{P}_1$ intercombination transition at 230.6\,nm is still accessible with standard laser systems and the narrow linewidth of 360\,kHz makes it an interesting candidate for laser-based high-voltage metrology \cite{Koenig.2020}. The transition is still strong enough to perform collinear fluorescence laser spectroscopy, but is to our knowledge  the slowest transition investigated with that technique so far.

At other facilities, collinear laser spectroscopy was performed over a wide range of indium isotopes, particularly to investigate nuclear properties close to the $Z=50$ shell closure \cite{Ulm.1985, Eberz.1987, Sahoo.2020}. However, in all cases atomic beams were used. Therefore, we will evaluate the accessibility of the ionic state for complementary measurements. This is of special interest for the hyperfine structure parameters $A$ and $B$, where disagreement was found between experiment and theory in recent studies of atomic $5p\rightarrow ns$ transitions \cite{Vernon.2020}. This indicates an increased importance of electron correlations in the excited atomic states. In the ion, the hyperfine interaction is about an order of magnitude stronger than in the atomic states and might therefore be a guide for improved coupled cluster calculations. 
So far, only a few ionic transitions have been measured with high precision. Among them, the forbidden $5s^2\;^1\mathrm{S}_0 \rightarrow 5s5p\;^3\mathrm{P}_0$ transition, which is employed for optical clocks \cite{Wang.2007b, Keller.2019}. In this context, also one hyperfine component ($F=\nicefrac{9}{2} \rightarrow F'=\nicefrac{11}{2}$) of the $5s^2\;^1\mathrm{S}_0 \rightarrow 5s5p\;^3\mathrm{P}_1$ intercombination  transition has been precisely measured as it is routinely employed as cooling transition \cite{Wang.2007}. The other components and, hence, the $A$ and $B$ factors are known with significant lower accuracies. We were able to improve those by two orders of magnitude.

\section{Experimental setup}
\label{sec:setup}
The measurements were conducted at the new Collinear Apparatus for Laser Spectroscopy and Applied Science (COALA) at TU Darmstadt. We identified the rest-frame transition frequencies $\nu_0$ of all three hyperfine transitions ($F=\nicefrac{9}{2} \rightarrow F'=\nicefrac{7}{2}, \,\nicefrac{9}{2}, \,\nicefrac{11}{2}$) of $^{115}$In$^+$ via collinear/anticollinear spectroscopy:
\begin{equation}
\nu_\textnormal{c} \cdot \nu_\textnormal{a} = \nu_0 \gamma (1+\beta) \cdot \nu_0 \gamma (1-\beta) = \nu_0^2 
\label{Eq:ColAcol}
\end{equation}
where $\nu_c$ and $\nu_a$ are the collinear and anticollinear laser frequencies at resonance with the Doppler-shifted transition frequency, $\beta$ is the ion velocity and $\gamma$ the time-dilation factor. A detailed description of the experimental realization including all applied ion sources, beamline and laser systems is given in \cite{Koenig.2020}.

\begin{figure}
	\centering
		\includegraphics[width=0.4800\textwidth]{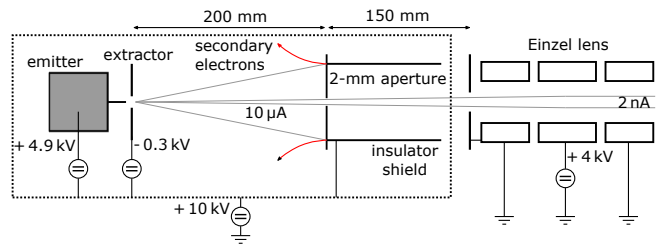}
	\caption{
	Adaption of a liquid-metal ion source for collinear laser spectroscopy. The ion current is kept constant by regulating the voltage applied to the emitter module while the total ion energy is stabilized by adapting the potential of the high voltage platform.
	It turned out that it is important to operate the trimming aperture on the same potential as the LMIS, since secondary electrons, generated by ion impingement, are otherwise accelerated back to the source. This deteriorates the ion-beam stability strongly \cite{Czarczyuski.1995,Mair.1997}. In our case, the relatively slow electrons can be deflected by floating the extractor plate on a negative potential of $-300$\,V. Behind the aperture a ceramic insulator connects the source to the beamline at ground potential. The ion beam is collimated by an Einzel lens.
	}
	\label{fig:LMIS-setup}
\end{figure}

Here, we will give a brief explanation with a focus on the operation of the liquid-metal ion source (LMIS) and the beam preparation, since to our knowledge such a source has not been used in collinear spectroscopy previously. 
As depicted in Fig.\,\ref{fig:LMIS-setup}, the LMIS and the associated power supplies are floated on a high-voltage platform from which the In$^+$ ion beam is emitted with a total kinetic energy of 15\,keV to the beamline at ground potential.
The LMIS was developed and assembled at TU Dresden \cite{Pilz.2017}. Between the emitter module, consisting of a capillary (50\,$\upmu$m bore diameter) on top of a heated reservoir filled with indium, and an extractor electrode in a few mm distance, a potential difference of about 5.2\,kV is applied.
The electric field is strong enough to ionize the liquid metal, which is fed through the capillary and the generated ions are then accelerated by the same field \cite{Orloff.2003, Forbest.1997}.
We stabilize the emitted ion current by operating the power supply floating the emitter module at a current limit of typically 10\,$\upmu$A. This evokes voltage fluctuations which we counteract by adapting the potential of the high-voltage platform, leading to an ion beam with constant current and ion energy. 

Even though the beam emitted from an LMIS has a large divergence \cite{Alton.1991,Loeffelmacher.1998}, it can be used to generate beams with very low transverse emittance, since it originates from a spot with diameter of less than 100\,nm. We restrict the beam to a 2-mm diameter aperture in a distance of 200\,mm from the emitter and estimate a normalized emittance of 10--50\,$\pi$\,mm\,mrad\,$\sqrt{\textnormal{eV}}$ by comparing the transmission through the pinhole with the divergence angles measured with a similar LMIS \cite{Koenig.2020, Vasiljevich.2011}. 
After trimming most of the ion beam, the remaining 2-nA beam passes a ceramic high-voltage break and is collimated by an Einzel lens in 150\,mm distance, finally providing a beam with a longitudinal spread of 22\,eV. 
Larger ion currents can be achieved by using larger apertures or larger emitter currents. We tested stable operation of up to 250\,$\upmu$A (before the aperture) that comes at the cost of a larger longitudinal energy spread \cite{Swanson.1983}.

The longitudinal velocity distribution leads to Doppler broadening which is significantly compressed by the acceleration to 15\,keV. Nevertheless, already a kinetic energy width of 22\,eV  causes a linewidth of about 500\,MHz which is much larger than the natural linewidth of 360\,kHz of the investigated intercombination transition.
This can be cured by using, e.g., gas-filled radio-frequency traps to cool the beam that are installed at many other laser spectroscopy facilities. As an alternative, we present a proof-of-principle realization of a pump-and-probe scheme that creates sharp spectral features on top of the Doppler broadened transition.

\begin{figure*}
	\centering
		\includegraphics[width=1\textwidth]{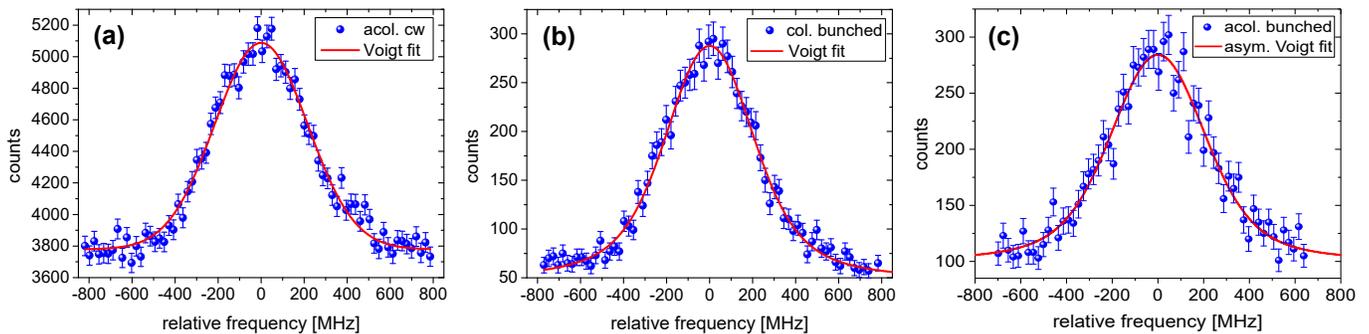}
	\caption{Resonance spectra of the $5s^2\;^1\mathrm{S}_0 \;F=\nicefrac{9}{2} \rightarrow 5s5p\;^3\mathrm{P}_1 \; F'= \nicefrac{11}{2}$ transition of ${}^{115}$In$^+$ obtained by varying the scan voltage applied to the drift tube in front of the fluorescence detection region (FDR). (a): Continuous measurement in anticollinear geometry. (b): The collinear laser is shuttered with an AOM while observing the fluorescence light in the FDR to remove the laser-induced background. In both cases the spectra are well described by a Voigt profile with a width of $\Gamma_\textnormal{FWHM} \approx  500$\,MHz, dominated by the Doppler width. (c): Laser-ion overlap was worsened to estimate its systematic impact (see text). To account for the resulting asymmetric lineshape, an asymmetric contribution has been included in the applied fit function \cite{Stancik.2008}.}
	\label{fig:Spectra}
\end{figure*}

After a $10\,^\circ$ bend, we superimpose the ion beam with the laser beams that are either copropagating (collinear) or counterpropagating (anticollinear) to the ion beam. After passing through ion optics for beam manipulation, an iris aperture with adjustable size localizes the position of all beams. In a distance of 2.6\,m downstream, a second iris is installed, hence defining angle and position of all beams.
Several arrays of drift tubes and a fluorescence detection region (FDR) that can be floated on individual potentials, are placed between the apertures. In particular, a first 1.2-m long drift tube enables optical manipulations of the ion beam. A second 0.17-m long drift tube is placed directly in front of the FDR and can be used for probing slow transitions. 
Our FDR is especially designed for spectroscopy in the deep UV-range and the combination of imaging and non-imaging optics leads to a passive background suppression \cite{Maass.2020}. Photons that are emitted from the ion beam are collected with a mirror array covering a solid angle of nearly $4\pi$ and counted with photomultiplier tubes.
The relatively long lifetime of 440\,ns of the excited state of the investigated $5s^2\;^1\mathrm{S}_0 \rightarrow 5s5p\;^3\mathrm{P}_1$ intercombination transition in combination with the fast ion beam that travels at a speed of 16\,cm/$\upmu$s allows us to probe the transition in the drift tube in front of the FDR, shutter the laser beam with an acousto-optical modulator (AOM) and monitor the fluorescence photons while the laser is blocked. This completely eliminates the otherwise dominant background from scattered laser light.

The laser system consists of two Sirah Matisse 2 TS Ti:Sapphire lasers, which are operated at 920\,nm and pumped with 19\,W of 532-nm light from two Spectra Physics Millennia eV lasers each. The light is quadrupled in four cavity-based frequency doublers (Spectra Physics, Wavetrain 2) yielding up to 100\,mW of 230-nm light. Both Ti:Sapphire lasers are locked to a Menlo Systems FC1500-250-WG frequency comb by a new locking scheme developed in cooperation with both manufacturers (see \cite{Koenig.2020} for more details).
The results presented here, are the first application of this locking scheme, achieving a stability of about 60\,kHz for the fundamental frequency at a time scale of minutes. Due to the direct lock to a frequency standard via the comb, long-term drifts are suppressed.
The UV light is sent through spatial filters to create collimated TEM$_{00}$ beams, which are then guided 5\,m through air to the downstream end of the beamline. While the anticollinear laser beam is directly transmitted through a plane uncoated quartz window into the beamline, the collinear laser beam is transported another 5.5\,m through air parallel to the beamline, before injecting the beam in opposite direction into the collinear apparatus through a similar window.


\section{Results}
\label{sec:results}
We investigated the intercombination transition $5s^2\;^1\mathrm{S}_0 \rightarrow 5s5p\;^3\mathrm{P}_1$ of In$^+$ in order to use this transition for high-voltage metrology \cite{Koenig.2020}. It consists of three hyperfine transitions ($F=\nicefrac{9}{2} \rightarrow F'=\nicefrac{7}{2},\,\nicefrac{9}{2},\,\nicefrac{11}{2})$ separated by about 35\,GHz, whereof the last one is the so-called cooling transition for optical clocks based on In$^+$.
As demonstrated in \cite{Koenig.2018}, we could clearly resolve both naturally abundant isotopes. However, we focused only on the more abundant $^{115}$In isotope (95.7\,\%) since we can deduce all relevant parameters for $^{113}$In from precise nuclear ratios measured in \cite{Rice.1957, Eck.1957a}.

While our setup is optimized for the anticollinear geometry, resulting in low background rates of only 10\,kHz for a laser power of 1\,mW, it is worse by a factor of 25 in the collinear case. 
As mentioned above, we used the drift tube in front of the fluorescence detection region for probing the ions. This was realized by Doppler tuning, i.e., the lasers are stabilized to a fixed frequency while the potential of the drift tube is scanned. The change of the ion velocity leads to varying Doppler shifts that are used to scan across the resonance. The fluorescence detection region was floated on a different potential to avoid laser-ion interactions outside the drift tube. We compared fluorescence detection under these conditions with direct excitation inside the FDR. According to the relative signal strength, 14\,\% of the photons are still detected when exciting the ion beam only inside the drift tube.
In Fig.\,\ref{fig:Spectra}a the optical response for the excitation of the In$^+$ beam in cw mode in anticollinear geometry inside the drift tube is shown. Due to the low background rate in this direction, a high signal-to-noise ratio of 22:1 is achieved within 8\,min of accumulation time. In Fig.\,\ref{fig:Spectra}b, a resonance spectrum taken in collinear geometry with a shuttered laser beam, is depicted. The beam-on/beam-off periods were chosen as $10\,\upmu$s/2\,$\upmu$s explaining the smaller signal strength for the same accumulation time. Due to the drastically reduced background, the signal-to-noise ratio is slightly improved to 25:1 despite the high background rates in this direction.
We identified the remaining background - unfortunately after the experiment - to be caused by the vacuum gauges. Turning those off will allow us to record practically background-free spectra in the future. 
In both cases, the resonances have a width of $\Gamma_\textnormal{FWHM} \approx 500$\,MHz and their resonance frequency can be determined with a statistical uncertainty of approximately 4\,MHz from the fit. The Voigt profile fitted to the experimental data is dominated by the Gaussian contribution which confirms the expectations on the ion production process regarding the initial velocity distribution.

For the precise determination of the three hyperfine transitions, we superimposed collinear and anticollinear lasers as well as the ion beam. The alignment of both laser beams could be easily checked in front of the anticollinear window where both beams arrive from the laser lab. Comparing the spot positions from one laser before and from the other after passing through the beamline and vice versa in a distance of 10\,m, enables an excellent superposition of both laser beams. Their size was adjusted to a diameter of 2\,mm to match with the ion beam size. By using the adjustable iris diaphragms inside the beamline, the ion-laser overlap was optimized.

For the determination of the rest-frame transition frequencies, a series of alternately performed collinear (c) and anticollinear (a) measurements was taken. To preclude uncertainties from unresolved long-term drifts, the applied measurement scheme was a-c-c-a.
As demonstrated in Eq.\,\ref{Eq:ColAcol}, the rest-frame transition frequency $\nu_0$ can be determined by multiplying the resonant frequencies $\nu_\textnormal{c}$ and $\nu_\textnormal{a}$ obtained in collinear and anticollinear measurements, respectively, if both are detected from an ion beam with identical velocity $\beta$. But since Doppler tuning is applied, the ion velocity is varied and this equation is not applicable as long as the two resonances do not appear at exactly the same scan voltage. Therefore, we chose both frequencies in a way that the resonance signal appeared at almost the same scan voltage. The remaining small deviation of the resonant scan voltage $\Delta U_\textnormal{scan}$ is corrected by linearly approximating its impact on the frequency of one laser
\begin{equation}
\delta \nu =
\frac{\partial \nu}{\partial U} \cdot \Delta U_\textnormal{scan} = \frac{2 \nu_\textnormal{0} q}{mc^2} \frac{\nu^2}{\nu^2-\nu_\textnormal{0}^2} \cdot \Delta  U_\textnormal{scan}
\label{eq:diffDoppler}
\end{equation} 
and including this in Eq.\,\ref{Eq:ColAcol}. This yields
\begin{equation}
\nu_0 = \sqrt{ \left( \nu_\textnormal{c} - \frac{\partial \nu_\textnormal{c}}{\partial U} \cdot \Delta U_\textnormal{scan} \right) \cdot \nu_\textnormal{a} } - \Delta \nu_\textnormal{recoil} ~.
\label{Eq:ColAcol2}
\end{equation}
Here, also the shift due to the photon recoil $\Delta \nu_\textnormal{recoil} \approx h\nu_0^2/2mc^2 = 33$\,kHz  is taken into account. Due to the long lifetime, we do not expect more than one interaction.
With this method the rest-frame transition frequencies between the ground state and all three hyperfine states of the $5s5p\,^3$P$_1$ level were measured. The results are listed in Tab.\,\ref{Tab:InColAcol}. By taking seven pairs of collinear and anticollinear measurements for the transition to the $F'=\nicefrac{11}{2}$ state and nine pairs for the transitions to the $F'=\nicefrac{9}{2}$ and $F'=\nicefrac{7}{2}$ states at the same day, the statistical uncertainty was reduced to 1.2\,MHz, 1.4\,MHz and 1.5\,MHz, respectively. This corresponds to the uncertainty of the weighted average that is slightly larger than the standard deviation of the mean. The systematic uncertainty is estimated to be less than 1.5\,MHz.

As demonstrated in \cite{Imgram.2019, Mueller.2020} it is generally possible to reduce the systematic uncertainty to 100\,kHz at COALA. However, the energy distribution of the ion beam used in those measurements was below 100\,meV compared to the width of about 22\,eV in our case. 
The dynamical properties of the electrostatic 10$\,^\circ$ deflector used to superimpose both beams, act as an energy filter and distribute ions with different kinetic energies in a direction perpendicular to the ion beam direction in the horizontal plane.
A misalignment between ion and laser beams can lead to a partial overlap of ion and laser beam and exclude a part of the velocity distribution from the interaction with the laser light.
If the two laser beams differ in position or diameter, this asymmetry might affect both spectra (a,c) differently and, hence, can cause uncorrelated shifts when extracting the centroid frequencies.
To quantify this uncertainty, a series of collinear/anticollinear measurements was performed with the overlap of both beams worsened on purpose. The spectra were fitted with an asymmetric profile \cite{Stancik.2008} which becomes necessary for a proper description of the data as shown exemplary in Fig.\,\ref{fig:Spectra}c.
Besides a higher statistical uncertainty of 2.2\,MHz, the center of the rest-frame transition frequency extracted from these measurements is shifted by 1.5\,MHz compared to the results listed in Tab.\,\ref{Tab:InColAcol}. Even though this shift is within the statistical uncertainty bands, it is conservatively considered as systematic contribution to account for possible deviations due to the asymmetric shape caused by a partial overlap of ion and laser beams.

\begin{table}
\centering
	\caption{Rest-frame transition frequencies and hyperfine parameters $A$ and $B$ of the $5s^2\,\,{}^1$S$_0 \rightarrow 5s5p\,\,{}^3$P$_1$ transition in $^{115}$In$^+$. 
Our values are in excellent agreement with ion-trap measurements \cite{Wang.2007, Peik.1994} but provide considerably improved accuracy in most cases. The measurement performed with a monochromator \cite{Larkins.1993} shows a systematic deviation of almost 300\,MHz, indicating a wrong calibration of their apparatus. This was also pointed out in a re-evaluation of the available literature data \cite{Kramida.2013}, in which a recalibrated value for the center-of-gravity frequency $\nu_\mathrm{cg}$ was published.
}
\begin{tabularx}{0.49\textwidth}{c l l r}
\addlinespace[.6em]
\hline
\addlinespace[.2em]
Transition	 &  \multicolumn{1}{c}{This work} & \multicolumn{1}{c}{Literature}  & Ref.\\ 
			 &  \multicolumn{1}{c}{MHz} & \multicolumn{1}{c}{MHz}  & \\ 
\addlinespace[.2em]
\hline
\addlinespace[.2em]

$\nicefrac{9}{2} \rightarrow \nicefrac{11}{2}$        & 1\,299\,648\,955.\,7\,(1.9) & 1\,299\,648\,954.\,54\,(10) & \cite{Wang.2007}\\
                            &                                    & 1\,299\,648\,690\,(150) & \cite{Larkins.1993}\\
$\nicefrac{9}{2} \rightarrow \nicefrac{9}{2}$       & 1\,299\,611\,097.\,9\,(2.1) & 1\,299\,610\,830\,(180) & \cite{Larkins.1993}\\
$\nicefrac{9}{2} \rightarrow \nicefrac{7}{2}$       & 1\,299\,579\,291.\,4\,(2.1) & 1\,299\,578\,990\,(180) & \cite{Larkins.1993}\\
$A$                         & ~\,~~\;~\,~~~~6\,957.\,19\,(28)\,    & ~\,~~\;~\,~~~~6\,960\,(90) &\cite{Peik.1994}\\
$B$                         &  ~\,~~\,~\,~~~~$-443$.\,7\,(2.4)   & ~\,~~\,~\,~~~~$-480$\,(420) &\cite{Peik.1994}\\
$\nu_\mathrm{cg}$           & 1\,299\,617\,759.\,3\,(1.2)        & 1\,299\,617\,480\,(150) &\cite{Larkins.1993}\\
                            &                                    & 1\,299\,617\,744\,(45) &\cite{Kramida.2013}\\
\addlinespace[.2em]
\hline
\end{tabularx}
\label{Tab:InColAcol}
\end{table}

\section{Discussion}
Three groups have previously reported measurements on the $5s^2\,\,{}^1$S$_0$\,$\rightarrow 5s5p\,\,{}^3$P$_1$ transition of $^{115}$In$^+$ neglecting the measurement of Paschen and Campbell carried out in 1938 \cite{Paschen.1938}. Most recently Wang \textit{et al.}\ \cite{Wang.2007} measured the cooling transition ($F=\nicefrac{9}{2} \rightarrow F'=\nicefrac{11}{2}$) on a single ion in a radio-frequency trap with an accuracy of 100\,kHz by using a frequency comb. Our value agrees well within our uncertainty bands, which confirms the accuracy of our measurement. 
Another measurement performed in a radio-frequency trap by Peik \textit{et al.}\ \cite{Peik.1994} provided all hyperfine transitions but is not listed in Tab.\,\ref{Tab:InColAcol} since they specified large uncertainties of 600\,MHz. Nevertheless, it should be noted that their values are in good agreement and differ by less than 90\,MHz to our results.
A smaller uncertainty of 150-180\,MHz is stated by Larkins and Hannaford \cite{Larkins.1993} who performed measurements using an échelle monochromator.
However, their values show a systematic deviation of about 280\,MHz to our results and to those from Wang and coworkers. A similar deviation of 277\,MHz has been observed in the clock transition ($5s^2\,\,{}^1$S$_0$\,$\rightarrow 5s5p\,\,{}^3$P$_0$) at 236.5\,nm \cite{Wang.2007b}.
As pointed out in detail by Kramida \cite{Kramida.2013}, who summarized and re-evaluated all spectroscopic data of In$^+$, the calibration of the monochromator and the conversion from the air wavelength into vacuum wavenumbers of \cite{Larkins.1993} can be strongly affected by underestimated systematic effects. Based on the accurate value of \cite{Wang.2007}, Kramida recalibrated these measurements and gave a value for the center of gravity of the $5s^2\,\,{}^1$S$_0$\,$\rightarrow 5s5p\,\,{}^3$P$_1$ transition \cite{Kramida.2013}.
From our values we can also deduce the center-of-gravity frequency $\nu_\mathrm{cg}$ as well as the hyperfine parameters $A$ and $B$. These have been calculated analytically from
\begin{equation}
\Delta \nu = \frac{1}{2}AC + \frac{B}{8}\frac{3C(C+1)-4I(I+1)J(J+1)}{I(2I-1)J(2J-1)}
\end{equation}
where $\Delta \nu$ denotes the frequency difference from the individual hyperfine transitions to the center of gravity and $C=F(F+1)-I(I+1)-J(J+1)$. Our results are included in Tab.\,\ref{Tab:InColAcol}.
Only Peik \textit{et al.}\ explicitly stated the hyperfine parameters in their work (see Tab.\,\ref{Tab:InColAcol}) but these can be deduced from Larkins and Hannaford's values as well ($A=6960\,(26)$\,MHz and $B=-460\,(190)$\,MHz \cite{Larkins.1993}). Both are in excellent agreement with our results.
Moreover, the re-evaluated center of gravity agrees nicely.

With the precise ratios of the magnetic dipole moment $\mu$ \cite{Rice.1957} and electric quadrupole moment $Q$ \cite{Eck.1957a} between $^{115}$In and $^{113}$In
\begin{equation*}
\begin{split}
\mu(^{115}\mathrm{In}^+)/\mu(^{113}\mathrm{In}^+)=1.0021437\,(12)  \\
Q(^{115}\mathrm{In}^+)/Q(^{113}\mathrm{In}^+)=1.0138236\,(13) 
\end{split}
\end{equation*}
we can calculate the hyperfine constants of $^{113}$In$^+$ to be $A(^{113}\mathrm{In}^+)=6942.32(75)$\,MHz and $B(^{113}\mathrm{In}^+)=-437.7\,(2.4)$\,MHz. We included an uncertainty contribution considering a hyperfine structure anomaly, which we estimate to be smaller than 0.01\,\% since both isotopes have the same nuclear spin of $I=\nicefrac{9}{2}$ and because measurements in the P$_{1/2}$ and P$_{3/2}$ states of atomic indium revealed anomalies of only $7.5\cdot10^{-6}$ and $23.8\cdot10^{-6}$, respectively \cite{Eck.1957b}.

The hyperfine constants can also be employed to extract the isotope shift $\nu(^{113}\mathrm{In})-\nu(^{115}\mathrm{In})=696.3(2)(3.1)$\,MHz from the precise measurement of Wang \textit{et al.}\ \cite{Wang.2007}. The first uncertainty value of 0.2\,MHz corresponds to the combined experimental uncertainty of both experiments while the large systematic contribution of 3.1\,MHz is an estimation of a possible hyperfine structure anomaly contribution, being neglected in \cite{Wang.2007}. Based on experimental uncertainties only, they present an uncertainty of 1.68\,MHz but since the same analysis is applied, this systematic contribution needs to be considered as well.

\begin{figure*}
	\centering
		\includegraphics[width=1\textwidth]{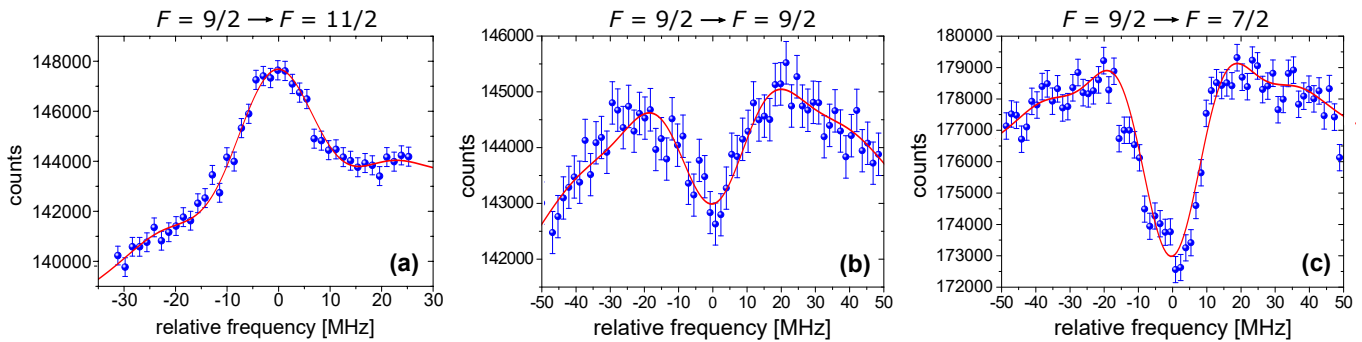}
	\caption{Spectral features imprinted on top of the main resonance of the different hyperfine transitions. By optically pumping one velocity class in the 1.2-m long drift tube in front of the fluorescence detection region (FDR), the population of the magnetic substates is altered, which is detected with the same laser by scanning the potential applied to the FDR. 
	In all cases, linearly polarized laser light is used. The direction of the features (peak or dip) and the relative intensity depends on the transition strengths of the different hyperfine transitions.
	}
	\label{fig:optPumping}
\end{figure*}

\section{Pump-and-Probe Measurements}

In order to overcome the limitations caused by the large velocity distribution inherent in the operation principle of the LMIS or of other ion source with relatively large energy spread, pump-and-probe measurements can be applied. For a proof of principle, we demonstrated this by operating the first 1.2-m long drift tube at a constant potential. One velocity class of the ion beam that is in resonance with the laser, will interact and can be manipulated. In the case of In$^+$, there is no hyperfine structure splitting in the $^1\mathrm{S}_0$ ground state. However, the hyperfine interaction still allows us to use laser light to reshuffle population between the $m_F=-\nicefrac{9}{2},\,-\nicefrac{7}{2},\, ...,\, +\nicefrac{9}{2}$ nuclear magnetic substates and to produce orientation or alignment of the nuclear spin. This property is conserved while the ion travels to the fluorescence detection region even though there is no guiding field as it is usually applied in collinear optical pumping measurements, as demonstrated for example in \cite{Arnold.1987, Voss.2013}. The reason is the long relaxation time caused by the ions' small magnetic moment being exclusively of nuclear origin, and the reduction of magnetic stray fields by proper shielding with permalloy. Subsequent excitation of the ions manipulated by the first laser interaction will lead to an increased or decreased fluorescence rate depending on the combination of laser polarizations that are used in the pump and the probe process. A narrow peak or a dip will appear accordingly on top of the Doppler broadened main resonance. 

We have realized this in a simple way using the same linearly-polarized light for both, optical pumping in the first drift tube and probing in the FDR. Therefore, the potential at the optical pumping region is chosen such that the laser beam is in resonance with ions close to the center of the Doppler broadened line. Then, the potential at the FDR is scanned to perform Doppler tuning around the center of the resonance. The peak or dip will only appear if ions of the same velocity class are addressed in both regions. The result is shown for all three hyperfine transitions in  Fig.\,\ref{fig:optPumping}. 

The sharp spectral feature observed in all three cases exhibits a width of about 14\,MHz on top of the main resonance. Please note that the $x$-axis spans only 60-100\,MHz compared to the more than 1.6-GHz wide region in Fig.\,\ref{fig:Spectra}. The width of the spectral feature is reduced by a factor of 35 compared to the Doppler-broadened peak but still considerably wider than the 360-kHz natural linewidth of the transition. A substantial fraction of this is expected to be caused by transit-time broadening and --  to a lesser extent -- saturation broadening. Transit-time broadening with a square-wave on/off-resonance condition which is given by the fast entrance and exit of the ions into the resonant potential inside the FDR, causes a $\sin^2(x)/x^2$-like lineshape. This function on top of a broad Gaussian has been used for the fits depicted in Fig.\,\ref{fig:optPumping} and is able to successfully reproduce the measured profile.
Nevertheless, based on the experimental conditions one would expect the resonance to be a factor of three narrower as it is observed. The reason for the additional broadening is not fully understood yet. 

On the other hand, the characteristic of the feature in terms of sign (dip or peak) and relative intensity agrees well with calculations based on the transition strength between the different $m_F$ states (Clebsch-Gordan coefficients) \cite{Koenig.2018}. The partial strength of  $\Delta m_F = \pm 1$ decays in the $F=\nicefrac{9}{2} \rightarrow F'=\nicefrac{11}{2}$ transition favors decay towards smaller $|m_F|$ values, where most of the $\pi$-transition strength is located. Therefore a peak is observed in Fig.\,\ref{fig:optPumping}a. While the preferred direction of decay is the same in the $F=\nicefrac{9}{2} \rightarrow F'=\nicefrac{9}{2}$ transition, the $\pi$-transition strength is concentrated on the transitions with large $|m_F|$ values. Hence, the total transition strength is weakened and a dip appears in Fig.\,\ref{fig:optPumping}b. Finally, in subfigure c, the dip is intuitively understandable since $\pi$-excitation will transport the population eventually into the $m_F=\pm \nicefrac{9}{2}$ states, which are dark states in the $F=\nicefrac{9}{2} \rightarrow F'=\nicefrac{7}{2}$ transition under these conditions.
This approach was successfully applied for high-voltage metrology measurements in \cite{Koenig.2020}. Furthermore, this linewidth reduction by a factor of 35 will allow even more precise determinations of the centroid frequencies and the systematic uncertainties due to a spatial energy separation will be avoided.

\section{Conclusion and Outlook}
We demonstrated that liquid-metal ion sources are well suited to generate metallic ion beams for high-precision collinear laser spectroscopy experiments. An excellent transverse emittance was achieved by trimming the beam. Despite the large longitudinal energy spread, we were able to present measurements of the three hyperfine transitions at the MHz level. Such measurements could further be improved, as demonstrated here, with the pump-and-probe technique.
The procedure is not limited to beams from LMIS but will be useful for all kind of beams from ion sources with large energy spread like, e.g., electron cyclotron resonance (ECR) sources and electron beam ion sources (EBIS). At COALA, we want to apply the latter to produce He-like ions of beryllium to carbon for laser spectroscopy in the $1s2s\; ^3\mathrm{S}_1 \rightarrow 1s2p\; ^3\mathrm{P}_J$ manifold. Combined with high-precision atomic structure theory \cite{Yerokhin.2018}, we expect to extract accurate nuclear charge radii directly from optical spectroscopy without the need to refer to electron scattering or muonic atom spectroscopy. Since the EBIS has a similar energy spread, pump-and-probe techniques will play a crucial role to reach the targeted accuracy of better than 1\,MHz. Depending on the element and nuclear spin of the isotope various techniques will be employed: saturation spectroscopy, $\Lambda$-spectroscopy or optical pumping as it was demonstrated here. 

\section{Acknowledgements}
We thank P. Laufer and M. Tajmar for their support during the commissioning of the LMIS.

We acknowledge support by the Deutsche Forschungsgemeinschaft (DFG, German Research Foundation) under Grant INST No. 163/392-1 FUGG and -- Projektnummer 279384907 -- SFB 1245, the Helmholtz International Center for FAIR (HIC4FAIR) and the Hessian Academy for FAIR (HFHF). K.K., P.I., B.M and T.R. acknowledge support from HGS-HIRE.

\bibliographystyle{aipnum4-1}
\bibliography{lit}

\end{document}